\documentclass[12pt]{iopart}

\usepackage{iopams}  
\usepackage{graphicx}
\graphicspath{ } 
\begin{document}

\title[Mirror-Bulk Interior of Schwarzschild BHs]{A Heuristic, Mirror-Bulk Model of Schwarzschild Black Hole Interiors}

\author{Noah M. MacKay
\\ ORCID ID: 0000-0001-6625-2321}

\address{Institut für Physik und Astronomie, Universit\"at Potsdam,\\ Karl-Liebknecht-Str. 24/25, 14476 Potsdam, Germany}
\ead{noah.mackay@uni-potsdam.de}
\vspace{10pt}
\begin{indented}
\item[]\today
\end{indented}

\begin{abstract}
Schwarzschild -- non-rotating and chargeless -- black holes (BHs) are classically understood to be voids of extreme gravitation, with modern and contemporary apporaches having involved string-theoretical black branes and superfluid analogs. In this study, we reconsider Schwarzschild BHs by letting its black brane be a closed spherical surface, within which a mirror-bulk fluid is contained, mirrored by surface-dwelling Hawking radiation quanta. By employing a once-used variational method of solving the Einstein field equations, the eneregtic output is approximated by the metric Ansatz input. Here, we consider a Tangherlini-like black 2-brane metric Ansatz in 3+1 spacetime, and via $G_{\mu\nu}=8\pi GT_{\mu\nu}$ the geometric information of the time-time component $G_{00}$ determines the corresponding (effective) energy density $T_{00}$ and the respective energy profile $E:=T_{00}V$. These are found to be functions of the radial coordinate, which also applies to the BH interior $r\in[0,~r_S]$. We explore and discuss the mirror-bulk energy content, internal pressure, and the mass- and pressure-gradients of a mirror-bulk BH.
\end{abstract}

%
%
%
\maketitle
%
%


\section{Introduction} \label{sec1}

Black holes (BHs), as described by general relativity (GR), are regions of extreme spacetime curvature where the event horizon at $r_S=2GM$ ($c=1$) and the geometric center impose mathematical singularities \cite{Schwarzschild:1916uq, Schwarzschild:1916ae, Penrose:1964wq}. Under a change of coordinates \cite{Eddington:1924pmh, Finkelstein:1958zz}, the mathematical singularity at the event horizon is removed, turning the event horizon into a passible region for inward moving observers. More contemporary approaches that utilize quantum field theory \cite{Hawking:1974rv, Hawking:1975vcx, York:1972sj, Gibbons:1976ue}, string theory \cite{Kovtun:2004de, Lunin:2001jy, Lunin:2002qf, Mathur:2002ie, Mathur:2005zp, Mathur:2008wi, Mathur:2008nj, Mathur:2009hf, Mathur:2012zp}, and viscous hydrodynamics \cite{Tangherlini:1963bw, Obers:2008pj} remove the event horizon coordinate in lieu of a non-trivial boundary manifold (i.e., the black brane). Implying both AdS/CFT and the holographic principle \cite{Kovtun:2004de, Lunin:2001jy}, the information encoded on (and emitted from) the black brane corresponds to a mirrored conformal bulk -- a holographic fluid -- below the manifold.

Previous ``mirror bulk" models of the BH interior include the Bose-Einstein condensate of gravitons \cite{Dvali:2011aa, Dvali:2012gb, Dvali:2012rt, Dvali:2014}, a bosonic superfluid \cite{Manikandan:2018urq}, and a thermal bath \cite{Hayden:2007cs}. The boson superfluid and thermal bath models were applied to late-stage evaporating BHs with a thermal noise-inducing temperature $T\propto1/M$. However, Hawking radiation emitted from a non-rotating horizon surface, hypothesized to consist 81\% neutrinos, 17\% photons, and 2\% gravitons \cite{Page:1976df}, enables one to propose another BH mirror bulk that is predominantly fermionic. And for astrophysical, (super-)massive BHs that are cold, a fermionic bulk would exhibit a Pauli degeneracy pressure and threshold Fermi energy. In other words, this gives a BH under this model ``internal structure", such as those within the stars they once were.  

As a means to investigate this heutistic and unconventional model, we employ a method analogous to quantum-mechanical variational method. Quantum-mechanical variational method estimates the energy eigenvalue of an impossible Hamiltonian by assuming an Ansatz wavefunction (one that is a known solution to the Schr\"odinger equation), computing the expectation value of the Hamiltonian, and minimizing it to approximate the energy at the ground state. In Refs.  \cite{MacKay:2024qxj, MacKay:2025uyg}, the Einstein field equations (EFEs) were solved using the similar philosophy: one begins by assuming a metric Ansatz that is well-defined under a choice of coordinates, and has a non-zero determinant, and apply differential operations that are inherent in the Christoffel symbols and the Ricci tensor to obtain effective energetic information. We adopt this philosophy in the present work; via $G_{\mu\nu}=8\pi GT_{\mu\nu}$ the geometric information of the time-time component determines the corresponding (effective) energy density $T_{00}$ and the respective energy $E:=T_{00}V$. In this work, our metric Ansatz is the Tangherlini-like black brane metric \cite{Tangherlini:1963bw, Obers:2008pj} for the event horizon surface, such that surface level energy (and if relevant within the radial span of the BH) can be discernible. We discuss this in Section \ref{metan}.

After reintroducing the variational approach to solving the EFEs in Section \ref{lbmeths}, we use this methodology to determine the effective energy density in Section \ref{engy}, and how this relates to the internal ``structure" of the mirror-bulk BH. Section \ref{discs} entails further discussion on BH internal pressure, along with heuristic arguments for the BH mass and pressure gradients and BHs in 5D (4+1) spacetime. Finally, we conclude in Section \ref{sect:concl} with a summary of findings and how this work may be expanded upon.  

\section{Methods}

\subsection{Metric of Choice: The Black Brane Metric} \label{metan}

Non-rotating BHs in GR conventionally follow the Schwarzschild metric. However, for a BH generally in $n$ physical dimensions, a BH itself can be expressed as a $p$-dimensional manifold: the Tangherlini-like black $p$-brane. This is described by the following metric \cite{Tangherlini:1963bw, Obers:2008pj}:
\begin{eqnarray}\nonumber
ds^2=\left(\eta_{\gamma\delta}+\frac{r_S^{n-p-3}}{r^{n-p-3}}u_\gamma u_\delta\right)d\sigma^\gamma d\sigma^\delta+\left(1- \frac{r_S^{n-p-3}}{r^{n-p-3}}\right)^{-1} &dr^2\\\label{tangh}
&+r^2d\Omega^2_{n-p-2},
\end{eqnarray} 
where $\eta_{\gamma\delta}$ is the $(p+1)$-Minkowski metric upon the black $p$-brane with the metric signature $(-,+\in\mathrm{dim}(p)\,)$. Transversely along the black brane, $u^\gamma=(1,\vec{0}_p)$ is the $(p+1)$-flow-velocity and $\sigma^\gamma=(t,\vec{x}_p)$ is a $(p+1)$-coordinate vector; $d\Omega^2_{n-p-2}$ defines the element of a $(n-p-2)$-sphere which the black brane encloses. 

For a point-like BH ($p=0$) upon $n=4$ physical spacetime, we recover the classic Schwarzschild metric. However, to approximate the BH event horizon as a a closed spherical surface, we must extend the dimensionality of the black brane, from the classic point-like BH to a two-dimensional surface ($p=2$). To maintain the intuition that this surface is within our 3+1 physical spacetime, we keep $n=4$. From this, one yields an unconventional metric of the following form: 
\begin{eqnarray} \nonumber
ds^2=-\left(1-\frac{r}{r_S}\right)&dt^2+\left(1-\frac{r}{r_S}\right)^{-1} dr^2+r^2\left(d\Omega^2_0+d\Omega^2_2\right)\\\label{2brane}
&+\sin^2\theta \left(dr^2-r^2 d\theta^2\right)+2r\sin\theta\cos\theta d\theta dr,
\end{eqnarray} 
which introduces peculiarities. For instance, $d\Omega^2_0=0$ for the 0-sphere (the disjoint union of two points: $S^0\simeq\, ^\bullet\coprod\,\! ^\bullet$) is obtained from the $d\Omega^2_{n-p-2}$ contribution, which otherwise suggests an absence of spherical symmetry. However, $d\Omega^2_2=d\theta^2+\sin^2\theta d\phi^2$ for the 2-sphere is preserved from the $d\sigma^\gamma d\sigma^\delta$ contribution, keeping spherical symmetry present in our metric, see footnote\footnote{For $p=2$, $\sigma^\mu=(t,x,y)\rightarrow(t,r\sin\theta\cos\phi,r\sin\theta\sin\phi)$ in the 3+1 spherical coordinate conversion.}. 

In addition to the leading metric that closely resembles the Schwarzschild metric, we also have a subleading metric only dependent on the radial $r$ and polar-angular $\theta$ coordinates. More specifically, this subleading metric has non-zero diagonal radial and polar terms that are commonly strong along the equator $\theta=\pi/2$ and vanishing at the poles $\theta=0,~\pi$. There are also non-zero radial-polar terms that resemble a form of off-diagonal shearing, which is at its extrema at the angles $\theta=\pi/4,~3\pi/4$: the angles between the equator and each pole. Given this and no diagonal azimuthal term in the subleading metric, these effects can be considered as anisotropic, as though perturbative. 

And so, one can define the subleading metric as a first-order perturbation metric, and we have the clear distinction in metric linearization in the form of $g+h$. One can therefore isolate the leading metric contribution to be our ``background" black 2-brane metric to analyze solely. In doing so, we must acknowledge nonetheless a present, perturbative structure which we set aside in this introductory work. 

For the main Schwarzschild-like contribution in Eq. (\ref{2brane}), this metric is more informative of the BH interior region, under the surface membrane. For $r<r_S$, Eq. (\ref{2brane}) reveals an internal linear-well profile with temporal and spatial flatness at $r=0$; the divergence is instead rimmed along the event horizon, at the coordinate singularity $r=r_S$. The idea of a linear-well profile inside a BH is unconventional yet previously considered under a different framework, i.e. Ref. \cite{MacKay:2025app}. However, it was used mainly as a mapping tool for outside observers to illustrate the linear downgrade of the BH horizon towards the geometric center, caused by continuous and uninterrupted Hawking radiation. Here, the mapping is treated as though literal yet with reversed mechanisms.

The divergence rimmed along the horizon acts heuristically as a gravitational sink, according to internal observers. One may alternatively reinterpret this as a ``repulsive" singularity, such that internal observers are not drawn inwards, but rather outwards while within the confines of the BH. Furthermore, if we consider a mirror bulk dictated by the Hawking particle microstates along the BH horizon seen from the outside (see Ref. \cite{MacKay:2025yja}), the mirror bulk particles inside the BH are drawn to the horizon, to eventually be radiated away into the vacuum as measurable Hawking radiation. 
 
However, outside the BH horizon as $r_S\ll r\rightarrow\infty$, Eq. (\ref{2brane}) yields a singularity-like diverence. This does not meet the convention of a flat spacetime infinitely far from the spherical body, which the conventional Schwarzschild metric captures. And according to Birkhoff's theorem, the metric of the external vacuum to any non-rotating object must be the Schwarzschild metric. To piece the external-vacuum Schwarzschild metric for $r>r_S$ with the Schwarzschild-like linear-well black brane metric for $r<r_S$, we have this piecewise metric:
\numparts 
\begin{equation}\label{newmet}
ds^2=-\mathcal{F}(r)dt^2+\frac{1}{\mathcal{F}(r)} dr^2+r^2d\Omega^2_2,
\end{equation}
where
\begin{equation}\label{pieces}
\mathcal{F}(r)=\left(1-\left(\frac{r_S}{r}\right)^{\Theta(r)}\right)\quad\mathrm{with}\quad\Theta(r)=\mathrm{sgn}\left(1-\frac{r_S}{r}\right).
\end{equation}
\endnumparts
Here, $\mathrm{sgn}(a)$ is the signum function of $a$, which is positive for $a>0$, negative for $a<0$, and zero for $a=0$. In the above metric, $\Theta(r)$ gauges the sign of the exponent, which weaves together the Schwarzschild exterior with the linear well interior. 

\subsubsection{Resting Points and their Stability}

By design, the event horizon at $r=r_S^\pm$ from both inside ($r=r_S^-$) and outside ($r=r_S^+$) is the common point of attraction. To verify this, we consider $\mathcal{F}(r)$ as a quintessential potential with points of stability. For feasibility, we impose the variable change $x:=r/r_S$ to have 
\begin{equation}
\mathcal{F}(x)=\left(1-x^{-\Theta(x)}\right)\quad\mathrm{with}\quad\Theta(x)=\mathrm{sgn}\left(1-\frac{1}{x}\right).
\end{equation}
By minimizing the potential $\mathcal{F}'(x)=0$, one obtains the resting points (i.e., the zeroes) $x_0$ of the original function:
 \begin{equation}
\mathcal{F}'(x)=-\frac{d}{dx}\left(x^{-\Theta(x)}\right)\Rightarrow x^{-\Theta(x)}\left[\Theta'(x)\ln(x)+\frac{\Theta(x)}{x} \right]=0.
\end{equation}
Being that the signum function has the similar mathematical functionality as the Heaviside function, its derivative is proportional to the Dirac delta: $\Theta'(x)=2\delta(x-1)$. The zeroes where $\mathcal{F}'(x)=0$ is satisfied are at $x_0=1^\pm$ (equivalently $r=r_S^\pm$) and as $x_0\rightarrow\infty^+$. Here, we only consider the fixed point of $x_0=1^\pm$.

For the BH's geometric center at $x=0$, where $\Theta(0)=-1$ and $\delta(-1)=0$, we obtain $\mathcal{F}'(0)=-1$. Because the first-derivative function is non-zero and negative at $x=0$, the BH singularity in this model is not a natural resting point. Instead, it has a negative gradient; any internal object near $x=0$ must be drawn away from this point, towards the closest resting point at $x_0=1^-$.

The second derivative, $\mathcal{F}''(x)$, tells us the curvature of the resting points just found. Negative curvature at $x_0$ resembles a true maximum of the original function: instability at $x_0$. Likewise, positive curvature at $x_0$ resembles a true minimum of the original function: stability at $x_0$. One can find that
 \begin{eqnarray}
\mathcal{F}''(x)\Rightarrow x^{-\Theta(x)}\Bigg[&-4\delta^2(x-1)\ln(x)^2-\frac{\Theta(x)^2}{x^2}-\frac{4\delta(x-1)\ln(x)}{x}\\\nonumber
&+\frac{4\delta(x-1)}{x}-2\delta(x-1)-\frac{\Theta(x)}{x^2} \Bigg],
\end{eqnarray}
where $\mathcal{F}''(x_0=1^\pm)=2$. In other words, the Schwarzschild radius of our mirror-bulk BH is a stable resting point, from both inside the BH and outside it.  

\begin{figure}[h!]
\centering
\includegraphics[width=0.65\textwidth]{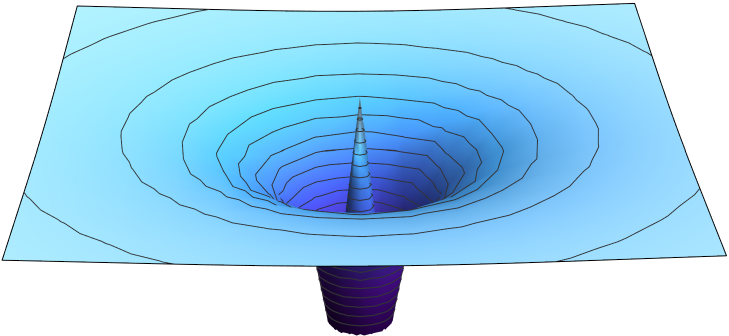}
\caption{\label{land} The spacetime manifold for a 2-brane Schwarzschild black hole as drawn by $\mathcal{F}(x)$ in Eq. (\ref{newmet}), with circular contour lines giving a cartographic representation of depth. The conic spike depicting the black hole linear well interior is a consequence to the black 2-brane, providing the intuition that the horizon is the attractive region.}
\end{figure}

For external observers, infalling objects will be suspended along the horizon, preserving conventional insight. For internal observers, objects inside the BH are drawn to the horizon under a linear well profile, to be collected along the horizon. It is, therefore, to be expected for co-moving observers that falling into the horizon would initially seem like gravitational free-fall, with horizon-rimmed attraction drawing the observer back to the horizon. This may repeat as though the co-moving observer were a damped harmonic oscillator along the horizon $x_0=1^\pm$, alternating between $x_0=1^+$ from the outside and $x_0=1^-$ from the inside. Such an occurance is reserved for a future study; nonetheless, what would be the emitted Hawking radiation from this model of BH is indeed the information of previous infalling objects huddled along the horizon. A visual aid of the spacetime landscape, plotted by $\mathcal{F}(x)$ in Eq. (\ref{newmet}), is provided in Figure \ref{land}.

\subsection{Variational Method of the EFEs}\label{lbmeths}

The Einstein field equations (EFEs) describe gravity as the spacetime curvature caused by energy-momentum sources:
\begin{equation}\label{efes}
R_{\mu\nu}-\frac{1}{2}Rg_{\mu\nu}=:G_{\mu\nu}=8\pi GT_{\mu\nu}.
\end{equation}
The left-hand side encodes the warped geometric response sourced by the energy-momentum source $T_{\mu\nu}$, in the shorthand known as the Einstein tensor $G_{\mu\nu}$. The Einstein tensor is linearly constructed from the metric $g_{\mu\nu}$ and Ricci curvature $R_{\mu\nu}$ tensors, and the Ricci trace scalar $R=g^{\mu\nu}R_{\mu\nu}$, where $g^{\mu\nu}$ is the inverse metric tensor. The additive cosmological term $\Lambda g_{\mu\nu}$ is neglected. 

Systematically, the EFEs are a collection of coupled, nonlinear equations. The geometry embedded in $G_{\mu\nu}$ depends on (besides the metric and Ricci tensors) relevant Christoffel symbols $\Gamma^\alpha_{\mu\nu}$, and one computes $G_{\mu\nu}$ generally by directly following a standard hierarchial method in solving the EFEs. Known solutions to the EFEs, e.g. Schwarzschild \cite{Schwarzschild:1916uq, Schwarzschild:1916ae}, Tolman-Oppenheimer-Volkoff (TOV) \cite{Tolman:1939jz, Oppenheimer:1939ne}, and Kerr \cite{Kerr:1963ud, Boyer:1967, Chandrasekhar:1985kt}, were obtained by first assigning a well-defined energy-momentum source for $T_{\mu\nu}$: either vacuum $T_{\mu\nu}=0$ \cite{Schwarzschild:1916uq, Schwarzschild:1916ae, Kerr:1963ud, Boyer:1967} or perfect fluid $T_{\mu\nu}=(\epsilon+P)u_\mu u_\nu+Pg_{\mu\nu}$ \cite{Tolman:1939jz, Oppenheimer:1939ne}. More generally, an exact analytical treatment of the EFEs becomes intractable for more complicated astrophysical systems. This has motivated several complementary approaches, such as the 3+1 decomposition in numerical relativity via the Arnowitt-Deser-Misner (ADM) formalism \cite{Arnowitt:1962hi} and heuristic analysis of spacelike singularities through the Belinksi-Khalatnikov-Lifshitz (BKL) conjecture \cite{Belinski:1971, Belinskii:1972sg, Belinskii:1973sud, BK, Belinsky:1982pk, Belinski:2017fas}.

The independent analysis by Chow and Knopf \cite{Chow:2004} (i.e., their lemma 3.32) suggests that, in appropriate coordinate settings, the Ricci tensor may be expressed schematically as the partial-differential Laplace-Beltrami operator acting on the metric, see footnote\footnote{Schematically, the Ricci tensor obeys $R\sim\partial\Gamma+\Gamma\Gamma$ and the Christoffel symbols satisfy $\Gamma\sim g^{-1}\partial g$. Thus, we claim that the Ricci tensor behaves qualitatively like a Laplacian operator acting on the metric tensor: $R\sim \nabla(g^{-1}\partial g)\sim\nabla^2 g$, where $\nabla\sim\partial+\Gamma$ is the covariant derivative.}, supplemented by lower-order terms:
\begin{equation}\label{ricci}
R_{\mu\nu}\simeq-\frac{1}{2}\Delta^\mathrm{LB} g_{\mu\nu}+\left(\mathrm{lower~order~terms}\right).
\end{equation} 
The operator $\Delta^\mathrm{LB}$ is the Laplace-Beltrami operator, and when the metric is well-defined and has a non-zero determinant, it can be written in a coordinate expression (i.e. in the Christoffel symbol-free form) as:
\begin{equation}\label{lbo}
\Delta^\mathrm{LB}=\frac{1}{\sqrt{-g}}\,\partial_\alpha\left(\sqrt{-g}g^{\alpha\beta}\partial_\beta \right),
\end{equation}
where $\sqrt{-g}=\sqrt{-\mathrm{det}(g_{\mu\nu})}$. This observation has motivated a Laplace-Beltrami-based reformulation of the EFEs, where the Ricci tensor is expressed at face-value by this operator. This notion was first explored in the context of coalescing compact-body binaries and radiated gravitational waves in Ref. \cite{MacKay:2024qxj, MacKay:2025uyg}, where the Ansatz metric used was the Kerr metric. In the context of this work, the geometric content of the theory is recast in terms of differential operators acting directly on Ansatz metric components, i.e. those presented in Eq. (\ref{newmet}). This, in turn, constructs component-wise the corresponding (effective) energy-momentum tensor that the metric  Ansatz approximately represents. One must be aware that such identifications are effective descriptions rather than as exact tensorial relations.

\subsubsection{Comments on the Laplace-Beltrami formalism}

Given the coordinate-dependence in the Laplace-Beltrami operator via Eq. (\ref{lbo}), the definitions of both the Ricci tensor via Eq. (\ref{ricci}) and the Ricci scalar heuristically constructed as $R\propto g^{\mu\nu}(\Delta^\mathrm{LB}g_{\mu\nu})$ ultimately depend on the choice of coordinate system. Specifically considering spherical coordinates, wherein Eq. (\ref{newmet}) is expressed, $g_{33}\propto\sin^2\theta$ leads to $g^{33}\propto\csc^2\theta$ that is present in the $g^{33}R_{33}$ contribution in the trace. In the Laplace-Beltrami formalism, i.e. as found in Ref. \cite{MacKay:2024qxj}, this contribution is divergent along the poles $\theta=0,\pi$, leading to the entire scalar value of $R$ to be divergent under full-angular integration. Thus, as demonstrated in Refs.  \cite{MacKay:2024qxj, MacKay:2025uyg}, one would turn to using effective Ricci scalars as a means of analytical completeness. These include e.g. the conventional Ricci scalar of the metric Ansatz ($R=0$ if vacuum), and a curvature surrogate in terms of the Kretschmann scalar $K=R_{\mu\nu\alpha\beta}R^{\mu\nu\alpha\beta}$ (as pursued in \cite{MacKay:2024qxj, MacKay:2025uyg}).

For Eq. (\ref{newmet}) is a Schwarzschild-Tangherlini-like vacuum metric, and also motivated out of heuristics, we adopt the simplistic route of utilizing the vacuum Ricci scalar $R=0$. Even if $R_{\mu\nu}=0$ conventionally for vacuum metrics, the second-order Laplace-Beltrami formalism allows for non-zero, \textit{effective} values in $R_{\mu\nu}$ (and therewith $G_{\mu\nu}$), provided the components of $g_{\mu\nu}$ are distinctly coordinate dependent and well-defined. This is particularly useful for approximating our desired quantites for our system of interest, constructing its effective energy-momentum tensor $T_{\mu\nu}$ via Eq. (\ref{efes}) component-wise. See Ref. \cite{MacKay:2026ofs} and Appendix A in Ref. \cite{MacKay:2024qxj} for how one retains the vacuum condition $R_{\mu\nu}=0$ in this formalism, and how the Bianchi identities are upheld.

\section{The Energy of the Mirror-Bulked BH}\label{engy}

Solving the EFEs conventionally typically begins with an assigned source to set as $T_{\mu\nu}$. In this variational approach to solve the EFEs, the choice of metric Ansatz effectively approximates the energy content by feeding metric components into the differential operations within Eq. (\ref{lbo}) -- much like its quantum-mechanical counterpart. For our BH that has an internal mirror bulk underneath a black 2-brane, we set Eq. (\ref{newmet}) to be our metric Ansatz. With this, we uniquely define the Laplace-Beltrami operator that in part defines the Ricci tensor in this context. 

For the metric laid out in Eq. (\ref{newmet}), its determinant is $\mathrm{det}(g)=-r^4\sin^2\theta$, and for a diagonal metric $g^{aa}=1/g_{aa}$ for $a\in[0,3]$. And since all components depend on $r$ and $\theta$, the Laplace-Beltrami operator under the black brane metric Ansatz follows a compact form:
\begin{equation}\label{lb2b}
\Delta^\mathrm{LB}_\mathrm{2-brane}=\frac{1}{r^2\sin\theta}\left[\sin\theta\partial_r\left(r^2\mathcal{F}(r)\partial_r\right)+\partial_\theta\left(\sin\theta \partial_\theta \right) \right].
\end{equation}
In this framework, the methodical chain to follow is $g_{\mu\nu}\rightarrow R_{\mu\nu}\sim\Delta^\mathrm{LB}g_{\mu\nu}\rightarrow R= g^{\mu\nu}R_{\mu\nu}$ (or uniquely define $R_\mathrm{eff}$) $\rightarrow G_{\mu\nu}\rightarrow T_{\mu\nu}$, where from Eq. (\ref{efes}) the energy content of the Ansatz geometric signature is defined straight-forwardly as
\begin{equation}
T_{\mu\nu}=\frac{1}{8\pi G}G_{\mu\nu}.
\end{equation}
To determine the energy of a mirror-bulked BH under Eq. (\ref{newmet}), one needs to approximate the effective energy density $T_{00}$, which is extracted from $G_{00}\sim R_{00}-Rg_{00}$ whereby $R_{00}\sim\Delta^\mathrm{LB}g_{00}$. To yield the energy, $E=T_{00}V$ within a spherical volume. Via Eq. (\ref{newmet}), $g_{00}=-\mathcal{F}(r)$, which given Eq. (\ref{pieces}) is a piecewise function dependent on regions of $r$ with respect to $r_S=2GM$. Furthermore, as $g_{00}$ only depends on $r$, only the radial-coordinate contribution of Eq. (\ref{lb2b}) is needed. 

The corresponding Ricci component for the energy density, $R_{00}$, via Eqs. (\ref{ricci}) and (\ref{lb2b}) is
\numparts
\begin{equation}\label{r00}
R_{00}=\frac{1}{2r^2}\partial_r\left(r^2\mathcal{F}(r)\mathcal{F}'(r) \right),
\end{equation} 
where $^\prime\equiv d/dr$, which depends on radial regions:
\begin{eqnarray}
& R_{00}=\frac{1}{2r_Sr}\left(3\frac{r}{r_S}-2 \right)\quad \mathrm{for}\quad r\leq r_S^-, \\
& R_{00}=\frac{r_S^2}{2r^4}\quad \mathrm{for}\quad r\geq r_S^+.
\end{eqnarray}
\endnumparts
For $r=r_S$ exactly, we have a duality in evaluation. The signum function collapses the expression to zero, yet each expression for the interior and exterior has a common non-zero evaluation: $1/(2r_s^2)$. Provided this duality in results, it provides physical intuition that an observer's measurement directly on the event horizon is impossible. However, an observer's measurement of the horizon scale from the external vacuum or heuristically from the interior is possible, and both measurements are identical.

Under the simplistic case of $R=0$, in accordance to the choice of metric Ansatz used, the Einstein tensor component $G_{00}$ is approximated by $G_{00}\approx R_{00}$ via Eq. (\ref{r00}). The energy density $T_{00}$ and associating energy $E=T_{00}V$ is straight forward however region dependent:
\numparts
\begin{eqnarray}\label{t00}
T_{00}\approx\frac{1}{8\pi G}R_{00}&=\frac{1}{16\pi Gr_Sr}\left(3\frac{r}{r_S}-2 \right)\quad \mathrm{for}\quad r\leq r_S^-, \\
&=\frac{1}{16\pi G}\frac{r_S^2}{r^4}\quad \mathrm{for}\quad r\geq r_S^+,
\end{eqnarray} 
\endnumparts
and, in terms of a variable change $r/r_S=x$ and the BH mass via $r_S=2GM$, we have region-dependent energy:
\numparts
\begin{eqnarray}\label{ein}
&E_\mathrm{in}(x) =\frac{M}{6}x^2\left(3x-2 \right)\quad \mathrm{for}\quad x\leq 1^-, \\\label{eout}
&E_\mathrm{out}(x) = \frac{M}{6}\frac{1}{x}\quad \mathrm{for}\quad x\geq 1^+.
\end{eqnarray}
\endnumparts
For the horizon scale of $x=1^\pm$ under the new variable change, $E_\mathrm{in}(1^-)=E_\mathrm{out}(1^+)=E_\mathrm{max}$ commonly in both regions, which is measured as $M/6$.

\subsection{Internal Energy}

One will notice that Eq. (\ref{ein}) contains information of interior energy that is rather curious. Firstly, besides the geometric center $x=0$, the other location of $x=r/r_S$ where energy vanishes is at $x=2/3$ (equivalently $r=2r_S/3$). Secondly, the energy is negative within $x\in[0,~2/3]$, and positive within $x\in[2/3,~1]$. 

It is the accepted understanding that negative energy, in the particle physics context, designates anti-particles, and we remind again that because of the holographic principle our mirror bulk is projected by the horizon-dwelling Hawking radiation particles. These compose of 81\% neutrinos, 17\% photons, and 2\% gravitons for non-rotating BHs \cite{Page:1976df}. Bosons (i.e. the photons and gravitons) are their own anti-particle, yet neutrinos of any lepton flavor (electron, muon, tau) are strictly ``standard" positive-energy particles, see footnote\footnote{That is, if we treat neutrinos as Dirac particles. For context, neutrinos as Majorana particles are their own anti-particle.}. 

From this knowledge, we can heuristically state that the mirror bulk BH has a negative energy ``outer core" that spans $x\in[0,~2/3]$ and a positive energy ``radiative zone" that spans $x\in[2/3,~1]$. An immediate motivation of the latter is that the BH's radiative zone (the layer closest to the event horizon surface) will continouously emit, namely, Hawking radiation that outside observers see as emitted from the horizon. Furthermore, we claim that the mirror bulk is a mixture of (anti-)particles that are segregated in their respective energy zones. 

It is therefore desired to determine the total energy content from these consituents; since $E_\mathrm{in}(x)$ is a function dependent on the radial scale $x=r/r_S$, we consider an average energy. However, knowing there are regions where the energy function is negative and where it is positive, we individually integrate Eq. (\ref{ein}) with respect to $x$ over each respective region, averaging them over these subranges. Thus, one would have a ``negative" average energy in the negative-energy zone:
\begin{equation}
\langle E_\mathrm{in}\rangle^-=\frac{1}{(2/3-0)}\int_0^{2/3}E_\mathrm{in}(x) dx=-\frac{M}{81},
\end{equation}
and a ``positive" average energy in the positive-energy zone:
\begin{equation}
\langle E_\mathrm{in}\rangle^+=\frac{1}{(1-2/3)}\int_{2/3}^1E_\mathrm{in}(x) dx=\frac{43}{648}M.
\end{equation}
This route is chosen over full-range averaging within $x\in[0,~1]$. With the negative area under the energy curve canceling out its  equal but opposite counterpart in the positive area under the curve, the full-range average energy evaluated would nonetheless be positive yet reduced: $\langle E_\mathrm{in}\rangle=M/72$. On the other hand, in deliberately breaking up the negative evaluation from the positive evaluation as separate integrations, we consider the explicit contributions of the negative energy (from anti-particles) and the positive energy (from standard particles) separately. An effective total average energy would thus be the sum of the magnitudes of each contribution:
\begin{equation}
\langle E_\mathrm{in}\rangle^\mathrm{ETA}=|\langle E_\mathrm{in}\rangle^+|+|\langle E_\mathrm{in}\rangle^-|=\frac{17}{216}M,
\end{equation}
which is less than $E_\mathrm{in}(1^-)=M/6$ by roughly one-half. From this addition, we can measure the percentages of each contribution: $|\langle E_\mathrm{in}\rangle^+|/\langle E_\mathrm{in}\rangle^\mathrm{ETA}=84.31\%$ for standard particles and $|\langle E_\mathrm{in}\rangle^-|/\langle E_\mathrm{in}\rangle^\mathrm{ETA}=15.69\%$ for anti-particles.

From these percentages, all 81\% of Hawking quanta that belong to neutrinos are in the positive-energy zone. Therefore, all bosons that make up the remaining 19\% of Hawking quanta entirely occupy the anti-particle zone and the rest of the positive energy zone. This is such that the mirror bulk of cold, astrophysical BHs would have a boson layer in the ``outer core" negative-energy zone and a portion of the ``radiative" positive-energy zone that is mainly fermionic. 

\subsubsection{Mirror-Bulk Fermi Energy}

To provide more support to this claim, we consider a back-of-the-envelope calculation of the Fermi energy $\varepsilon_F$. For thermal fermionic systems, the Fermi energy is a threshold energy at which fermions remain degenerate in their lowest occupied energy states (if $\varepsilon_F>k_BT$) or can advance into higher unoccupied states (if $\varepsilon_F<k_BT$). We consider the Fermi energy simply because neutrinos hold the majority among the emitted Hawking radiation / mirror bulk quanta. Bosons, on the other hand, are not bound by the Pauli exclusion principle and do not have an equivalent to Fermi energy. However, under the basis that the emission of Hawking quanta, no matter the particle type, behaviorally follows Fermi-Dirac statistics \cite{MacKay:2025yja}, our bosons will effectively behave like fermions. And so, we effectively treat our mixed-species mirror bulk of neutrinos and bosons as a ``pure" fermionic system.

For BHs, $k_BT=\hbar/(8\pi GM)$ using the Hawking temperature; $\hbar$ is the reduced Planck constant from quantum mechanics. The Fermi energy for massless particles (assuming neutrinos are massless):
\begin{equation}
\varepsilon_F=\hbar\left(3\pi^2\frac{N}{V}\right)^{1/3},
\end{equation}
depends on the total number of fermions $N$ and the volume $V$ of our enclosing. For our mirror-bulk particles, we consider the total number of Hawking quanta along the horizon (according to outside observers; found in Ref. \cite{MacKay:2025yja} to be $N=GM^2/\hbar$) and the effective spherical volume of a BH $V=4\pi r_S^3/3$. This yields, in terms of BH thermodynamic quantities:
\begin{equation}
\varepsilon_F=\frac{\hbar}{2GM}\left(\frac{9\pi r_S^2}{16 l_P^2}\right)^{1/3}\equiv 4\pi k_BT \left(\frac{9}{16}\frac{S}{k_B}\right)^{1/3}.
\end{equation}
Here, $S=k_BA/(4l_P^2)$ is the Hawking-Bekenstein formula for BH entropy ($l_P=\sqrt{\hbar G}\sim10^{-35}~\mathrm{m}$ is the Planck length; $2l_P$ is the Schwarzschild radius of a quantum BH \cite{Hawking:1971ei, Carr:2005qn}), where $A=4\pi r_S^2$ is the BH surface area. Knowing the Boltzmann formula for entropy is $S=k_B\ln\Omega$, where $\Omega$ is the multiplicity of microstates, the Fermi energy of our fermionic(-like) BH mirror bulk depends on the natural logarithm of this multiplicity, which is well-understood to scale in the same order of magnitude as the number of particles in the system: $\ln\Omega=A/(4l_P^2)=4\pi GM^2/\hbar\equiv 4\pi N$.

In comparison to the BH thermal energy $k_BT=\hbar/(8\pi GM)$, we verify that $\varepsilon_F>k_BT$ even if $\varepsilon_F\propto k_BT$. This is especially so for (super-)massive astrophysical BHs where $\varepsilon_F/k_BT\sim N^{1/3}\sim M^{2/3}$. This defines the mirror bulk as a degenerate medium, relying on the Pauli exclusion principle to give the BH interior a ``structure'' analogous to typical degenerate systems such as compact stars.

\subsubsection{Astrophysical Implication}

If we accept this proposition, early-stage Hawking radiation of a BH with initial mass $M_0$ is neutrino-rich. This is as long as the neutrinos mainly occupy the positive-energy zone closest to the BH horizon. Once the BH reaches a mass around $M\simeq2M_0/3$ -- assuming it hasn't taken in any more mass --, and knowing well that the BH radius scales with mass, the collective boson-rich layer is exposed, and the Hawking radiation from there on is bosonic and progressively thermal as the BH continues to evaporate. This idea furthermore gives support to the previous mirror bulk models that focus primarily on a boson condensate / superfluid medium \cite{Dvali:2011aa, Dvali:2012gb, Dvali:2012rt, Dvali:2014, Manikandan:2018urq}, which applies to BHs whose mirror-bulk is reflected by late-stage Hawking evaporation.

Recent literature suggest that neutrino emission as a form of Hawking radiation is sourced from evaporating primordial BHs \cite{Lysyy:2024qrs, Klipfel:2025jql, Anchordoqui:2025opy}. As primordial BHs are hypothesized to have formed just after the Big Bang, their neutrino-rich radiative layer will surely have already been radiated away by the present time. Under the basis that primordial BHs had emitted neutrinos completely as Hawking radiation, and given our model in which the neutrino-heavy positive-energy zone extends from $r\in[2/3,~1]r_S$, it is possible to estimate the mass of a primodial BH using the BH evaporation timescale \cite{Pickover:1996, Carroll:1996}, given by 
\begin{equation}\label{evap}
\tau_\mathrm{evap}(M_0)=\frac{5120\pi G^2M_0^3}{\hbar}
\end{equation}
for the initial black hole mass. For the difference in timescale $\Delta\tau_\mathrm{evap}=\tau_\mathrm{evap}(M_0)-\tau_\mathrm{evap}(2M_0/3)$, we yield
\begin{equation}\label{diff}
\Delta\tau_\mathrm{evap}=\frac{5120\pi G^2M_0^3}{\hbar}\left(\frac{19}{27} \right),
\end{equation}
which we set approximately equal to the current age of the universe at $t_\mathrm{Age}=13.8$ GYr. Setting $t_\mathrm{Age}\simeq\Delta\tau_\mathrm{evap}$ via Eq. (\ref{diff}), we estimate the mass of a primordial BH to be the real and positive answer of $M_0\simeq1.913\times10^{11}$ kg. 

On the other hand, in the case where all of the primordial BH mass is evaporated within the age of the universe, one sets $t_\mathrm{Age}=\tau_\mathrm{evap}(M_0)-\tau_\mathrm{evap}(0)$ via Eq. (\ref{evap}) to yield $M_0=1.702\times10^{11}$ kg. Therefore, following the premises of our model and what is suggested by recent literature \cite{Lysyy:2024qrs, Klipfel:2025jql, Anchordoqui:2025opy}, a primordial BH that has an interior mirror-bulk of our design would have the remaining mass of $M=2.11\times10^{10}$ kg, and the remaining lifetime of $26.317$ MYr. This should set possible motivation to detect such a primordial BH, which would from now on (if not taking in additional mass) only radiate photons and gravitons. Respectively, this would be in the form of blackbody radiation and gravitational radiation, although the nature of the latter is inconceivable without a quadrupolar structure.

On the other hand, astrophysical BHs whose initial mass (greatly) exceeds $3M_\odot$ ($M_\odot=2\times10^{30}$ kg defines the solar mass) would naturally emit neutrinos as Hawking radiation for as long as Eq. (\ref{diff}) given $M_0\geq3M_\odot$. That is, albeit tangent to the emission of neutrinos from astrophysical BHs via e.g. their jets.

\section{Discussion}\label{discs}

Following the holographic principle, horizon-dwelling Hawking particle microstates project the presence of a mirror bulk underneath the black brane surface. Given factors such as a cold temperature for astrophysical BHs via $T\propto1/M$, the emitted Hawking particles from such BHs are quantum-mechanically driven rather than thermally induced. This radiation is neutrino-heavy, and particle emission follows Fermi-Dirac statistics regardless of particle type \cite{MacKay:2025yja}. As a result, the mirror bulk the radiation reflects is expected to inherit a Pauli degenerate pressure that introduces internal structure to the BH. 

Intuitively, such an internal structure would exhibit an energy spectrum across the radial span of the BH. And via a method that is independent of any contemporary quantum-gravity or string-theoretical principles, such an energy spectrum is obtained as Eq. (\ref{ein}). This function for the internal energy is obtained by the variational approach of solving the EFEs, using the Laplace-Beltrami formalism for the Ricci tensor via Eq. (\ref{ricci}) and Eq. (\ref{newmet}) as our metric Ansatz.

\subsection{BH Internal Pressure}

Due to a present Hawking temperature $T=\hbar/(8\pi GM)$, BHs are essentially thermal systems. Under our model of a BH entailing a mirror bulk, each particle within this bulk are subject to thermodynamic principles. And thus, the mirror bulk BH would contain an effective pressure, which we simply obtain from the ideal gas law:
\begin{equation}\label{ideal}
P=\frac{N}{V}k_BT=\frac{3}{256\pi^2 G^3M^2}.
\end{equation}
Here, we recycle the number of particles $N=GM^2/\hbar$ and the effective volume of the BH $V=4\pi r_S^3/3$. With the absence of Planck's reduced constant, the BH pressure is not quantum mechanical. However, unlike the Oppenheimer-Snyder picture of a BH as dust-like \cite{Oppenheimer:1939ue}, the pressure is non-zero yet dependent on $1/M^2$. For a very large BH mass, the pressure is indeed miniscule, enough so to truncate it as zero for practical purposes.

One may find pressure alternatively via the Laplace-Beltrami formalism used to find the energy density. For e.g. the effective radial-radial stress component $T_{11}$, one computes the effective $G_{11}\approx R_{11}$ component, which via Eq. (\ref{ricci}) $R_{11}\sim\Delta^\mathrm{LB}g_{11}$ and via Eq. (\ref{newmet}) $g_{11}=1/\mathcal{F}(r)$. We only consider the internal profile, where $\Theta(r)=-1$, and under the variable change $x=r/r_S$ the interior region spans from $x\in[0,1]$. This gauges $x$ to be small, and one can impose a Taylor expansion on $g_{11}=1/\mathcal{F}(x)$ up to second order in $x$; this serves as a means to regulate the function:
\begin{equation}
g_{11,\mathrm{in}}\approx 1+x+\mathcal{O}(x^2)=1+\frac{r}{r_S}+\mathcal{O}(r^2).
\end{equation}
Defining $R_{11}\sim\Delta^\mathrm{LB}g_{11}$ for the interior, where $\mathcal{F}(r)=1-r/r_S$, comes to be
\begin{equation}
R_{11}\approx\frac{1}{2r^2}\left(3\frac{r^2}{r_S^2}-2\frac{r}{r_S} \right),
\end{equation}
and therefore $T_{11}\approx R_{11}/(8\pi G)$ for $R=0$. In terms of the variable change $x=r/r_S$ and $r_S=2GM$,
\begin{equation}\label{t11}
T_{11}\approx\frac{1}{64\pi G^3M^2}\frac{\left(3x^2-2x \right)}{x^2}.
\end{equation}
The radial-radial stress component, in short the radial pressure, is also a function of the radial span. At the geometric center at $x=0$, the radial pressure is infinite and positive. This intuitively defines an outward pressure otherwise present in typical stars, but this is enormously and rather unphysically large. Along the horizon scale $x=1^-$, the radial pressure has a convergent, positive value of $T_{11}=1/(64\pi G^3M^2)$. This has the consistent parameterization of $1/(G^3M^2)$ as in the ideal gas law alternative, with this route producing a pressure that is greater by a factor of $4\pi/3$ (i.e., by the volume of a unit sphere).

\subsubsection{Perturbative Radial Pressure}

Should one recall Eq. (\ref{2brane}), the leading internal linear-well metric has accompanying, subleading metric elements that we effectively dismissed as perturbative. This includes a secondary radial-radial component of $h_{11}=\sin^2\theta$, which heuristically produces a perturbative radial pressure in the form of $\delta T_{11}$. This is determined by the corresponding $\delta G_{11}\approx\delta R_{11}$, and we claim that $\delta R_{11}\sim \Delta^\mathrm{LB}h_{11}$ where the Laplace-Beltrami operator is defined only by the background metric (i.e. we imply Eq. [\ref{lb2b}]). This is so that we can decipher the perturbative information within the basis of our background metric, as one does for e.g. gravitational waves and tidal perturbations.

Provided $h_{11}$ is purely in terms of the polar angle, the polar-angular part of Eq. (\ref{lb2b}) shall be used, applying $\Theta(r)=-1$ to limit ourselves in the interior:
\begin{equation}
\delta R_{11}=-\frac{1}{2}\Delta^\mathrm{LB}_\mathrm{2-brane}h_{11}=-\frac{1}{2r^2}\left(1+3\cos(2\theta) \right).
\end{equation}
Because $\delta R_{11}$ is dependent on $\theta$, we must define this quantity for the whole spherical structure of the mirror-bulk BH. Thus, we integrate over all available polar angles $\theta\in[0,~\pi]$:
\begin{equation}
\Rightarrow \delta R_{11}=-\frac{\pi}{2r^2}.
\end{equation}
We use the above to define $\delta T_{11}\approx\delta R_{11}/(8\pi G)$ with $R=0$, and we define it specifically at $r=r_S$:
\begin{equation}
\delta T_{11}\approx\left.\frac{1}{8\pi G} \delta R_{11}\right|_{r=r_S}=-\frac{1}{256 G^3M^2}.
\end{equation}
This perturbative radial pressure, unlike Eq. (\ref{t11}), is negative in value. Therefore, the direction it points is radially inward and one can find a total radial pressure of $P_r=T_{11}+\delta T_{11}$ at $r=r_S$:
\begin{equation}
P_r=\frac{1}{64\pi G^3M^2}-\frac{1}{256 G^3M^2}=\frac{(4-\pi)}{256\pi G^3M^2}.
\end{equation}
This is still a positive-valued, outward-pointing pressure, yet it is reduced. If the radial pressure $T_{11}$ were instead the pressure found from the ideal gas law alternative, then we would have:
\begin{equation}\label{totpress}
P_r=\frac{3}{256\pi^2 G^3M^2}-\frac{1}{256 G^3M^2}=\frac{(3-\pi^2)}{256\pi^2 G^3M^2};
\end{equation}
this is, however, a negative-valued, inward-pointing pressure.

If one uses the ideal gas alternative for the BH pressure and the Laplace-Beltrami calculation for the perturbative pressure, the negative total pressure denotes the overpowering nature of a mechanism analogous to inward collapse. Although we found that the Schwarzschild radius is a stable resting point for our choice of metric Ansatz (Eq. [\ref{newmet}]), we recall that Hawking radiation is a radiative mechanism that causes the downgrade of the horizon inward. As long as the BH does not take in any more mass, its horizon scale will continue to decrease due to a net inward pressure, such as Eq. (\ref{totpress}).

\subsection{BH mass and pressure gradients}

With the ideal-gas pressure, it is possible to extract a pressure gradient $P'$ ($^\prime=d/dr$), and with it a mass gradient $m'$, for a mirror-bulk BH. While these usually apply to stars with hydrodynamic equilibrium, and certain expressions for the mass and pressure distributions depend on the star's equation of state, the mirror-bulk inside the BH can be approximated as a fluid or gas that obeys respectively hydrodynamics or statistical mechanics. Although we claim that this mirror-bulk consists of 81\% neutrinos, 17\% photons and 2\% gravitons, and comprehensive equations of state rely on microphysical processes, we take an atheistic approach that is independent of the mirror bulk composition. I.e., we present preliminary expressions for the BH pressure- and mass-gradients, if we allow Hawking radiation to dictate BH thermodynamics and with that the mirror-bulk's dynamics.

From Eq. (\ref{ideal}), where $P\propto M^{-2}$, let us treat these quantities as radial-coordinate dependent: $P\rightarrow P(r)$ and $M\rightarrow m(r)$. This way, we see that pressure changes with the mass, which itself changes with its radial span. Thus, we can straight forwardly assess, given $r_S=2GM$ at the surface horizon, that effectively
\begin{equation}
m(r)=\frac{r}{2G}\quad\Rightarrow\quad m'=\frac{1}{2G}.
\end{equation}
From this mass distribution, mass scales with radius $m(r)\sim r$; this is well-known for BHs. At the surface $r=r_S$, we retreive surface mass $m(r_S)=M$; at the geometric center we retrieve effectively no mass $m(0)=0$. These ``core" and ``surface" criteria for the mass are analogous to those for stars. One can, alternatively, use Eq. (\ref{ein}) and the mass-energy relation $E=M$ to readily define a mass distribution function $m(x)$, whose differentiation with respect to $x$ offers a spectral function in the mass gradient across the internal radial span.

With $m(r)=r/(2G)$, the pressure $P(r)$ and its gradient are found to be
\begin{equation}\label{presgrad}
P(r)=\frac{3}{64\pi^2 Gr^2}\quad\Rightarrow\quad P'=-\frac{3}{32\pi^2 Gr^3}=-\frac{2}{r}P(r).
\end{equation}
As it is intuitive in the mass-dependent case, the change in pressure with respect to the radial coordinate is negative. In other words, the change in pressure is inward-seeking, in alignment with the fact that Hawking radiation decreases the BH horizon scale, which increases $P(r)$. One can alternatively use Eq. (\ref{t11}) to define the radial pressure as a function of $x$, whose derivative with respect to $x$ defines the pressure gradient.

A stellar object's pressure is typically non-zero ar the core $r\rightarrow0$, defining a core density $P(0)=P_0$. However, we see that $r\rightarrow0$ causes the pressure to diverge, referring to Eq. (\ref{presgrad}). Nonetheless, we can take advantage of the fact that Hawking radiation, effectively, stops once the original BH becomes a quantum BH, with known mass $M_\mathrm{QBH}=\sqrt{\hbar/G}$ and radius $r_\mathrm{QBH}=2l_P$. We can therefore define the quantum BH pressure as a quintessential core pressure:
\begin{equation}
P(2l_P)=P_0=\frac{3}{64\pi^2 G^2\hbar},
\end{equation}
and we reassure that $m(2l_P)=l_P/G=\sqrt{\hbar/G}$, recovering the quantum BH mass.

\subsubsection{BH density and ``equation of state"}

It is common for stars and other hydrodynamic bodies to relate pressure in terms of mass/energy density, among other hydrodynamic/thermal quantities. We can heuristically extend this for our mirror-bulk BH. From $m(r)=\epsilon (r) V(r)$, we define simply:
\begin{equation}
\epsilon(r)=\frac{m(r)}{V(r)}=\frac{3}{8\pi Gr^2},
\end{equation}
which imposes a couple options to explore: (i) relate $\epsilon(r)$ with $P(r)$ to define $P\sim\epsilon$, or (ii) relate $\epsilon(r)$ with the effective energy density $T_{00}$ via Eq. (\ref{t00}) for the BH interior.

\begin{enumerate}
\item Following the option of $\epsilon(r)\sim P(r)$, we impose a basic proportionality $P=\eta \epsilon$, where one straight-forwardly finds the coefficient $\eta$:
\begin{equation}
\eta=\frac{P(r)}{\epsilon(r)}=\frac{1}{8\pi}\quad\Rightarrow\quad P=\frac{\epsilon}{8\pi}.
\end{equation}
One furthermore finds the square of the speed of sound by $c_S^2=dP/d\epsilon$, which is $1/(8\pi)$. Therefore, $c_S=1/\sqrt{8\pi}$, imposing that the mirror-bulk medium does not violate the causal limit of $c_S^2=1$.\\

We note that the pressure-density expression $P\sim\epsilon$ is typically obtained by the traceless condition of an (effective) energy-momentum tensor, assuming it follows an ideal/perfect fluid form:
\numparts
\begin{eqnarray}
&\epsilon-8\pi P=0=:g^{\mu\nu}T_{\mu\nu}\\
&\Rightarrow\,T^\mathrm{eff}_{\mu\nu}=\left(\epsilon+\frac{8\pi}{3}P\right)u_\mu u_\nu+\frac{8\pi}{3}Pg_{\mu\nu}.
\end{eqnarray} 
\endnumparts
While this is not the conventional perfect fluid energy-momentum tensor of form $T_{\mu\nu}=(\epsilon+P)u_\mu u_\nu+Pg_{\mu\nu}$, we see nonetheless that the mirror-bulk material inside our model of BH can approximate a near-pefect fluid. If we enforce $T_{\mu\nu}=(\epsilon+P)u_\mu u_\nu+Pg_{\mu\nu}$ to be the energy-momentum tensor at equilibrium $T^{(0)}_{\mu\nu}$, even for the mirror-bulk BH, the excess contribution of pressure $P$ heuristically describes a main-diagonal bulk-viscous tensor $B_{\mu\nu}$:
\begin{equation}
B_{\mu\nu}=\frac{(8\pi-3)}{3}P\left(u_\mu u_\nu+g_{\mu\nu}\right).
\end{equation} 
We recognize that, for a flat metric, $u_\mu u_\nu+g_{\mu\nu}=-\delta_{ij}$ for the 3-spatial dimensions, and typical bulk viscous tensors in the flat background follow the form $B_{ij}=\lambda\left(\vec{\nabla}\cdot\mathbf{u}\right)\delta_{ij}$, with $\lambda$ being the bulk viscosity. Therefore, if one makes the oversimplifying and naive assumption that the mirror-bulk material is flat everywhere, one has the following bulk-viscous relation:
\begin{equation}\label{bulky}
\lambda\left(\vec{\nabla}\cdot\mathbf{u}\right)=-\frac{(8\pi-3)}{3}P.
\end{equation} 
Optimistically speaking, this can be generalized for curved backgrounds, such as one proposed by Eq. (\ref{newmet}) for the BH interior.\\

\item Following the other option of $\epsilon(r)$ in relation to the interior $T_{00}$, we rewrite $T_{00}(r\leq r_S)$ with the revision $r_S\rightarrow r$ for a variable horizon scale:
\begin{equation}
\Rightarrow~T_{00}(r)=\frac{1}{16\pi Gr^2}=\frac{m'}{2A(r)},
\end{equation}
where $A(r)=4\pi r^2$ is the surface area element. This shows that $T_{00}\propto\epsilon(r)$ by a factor of $1/6$. We remind that Eq. (\ref{ein}), the energy distribution of the mirror-bulk interior, scales to $M/6$ at the horizon scale. This heuristically finds a flux in energy density to be a constant:
\begin{equation}
T_{00}(r)A(r)=\frac{m'}{2}.
\end{equation}
\end{enumerate}
 
\noindent And just as stellar objects have a core pressure, there is an associating core density:
\begin{equation}
\epsilon(2l_P)=\frac{3}{8\pi G^2\hbar}=\frac{6m'}{8\pi G\hbar}.
\end{equation}
One may notice that $8\pi G\hbar$, an area scale obtained in loop quantum gravity computations \cite{Rovelli:1987df, Rovelli:1989za, Rovelli:1994ge, Rovelli:2007}, is recovered.

\subsubsection{Future Study Topics}

It is of interest to take into account the proper microphysical processes that are relevant to the mirror-bulk consituents: the neutrinos of all flavors, photons and gravitons. It is well understood that neutrinos of one flavor can interact with those of the same flavor, just as they interact with those of a different flavor. Photons are not self-interacting, nor they interact with any particle except for the electron in conventional Compton scattering. One can suggest a Compton-like interaction between neutrinos and photons, and that between neutrinos and gravitons. Gravitons can self-interact, albeit graviton-graviton diagrams with internal loops notoriously lead to infinities not even renormalization can resolve.

The process of two gravitons annihilating into two photons (and the reverse process) have been explored previously \cite{Skobelev:1975, Blas:2026hao}, with more recent applications being the stochastic gravitational wave backround and binary mergers \cite{Blas:2026hao}. Such rules in the context of a mirror-bulk, and deducing which processes are allowed between neutrinos, photons, and gravitons, are reserved for a future study.

\subsection{BHs in $n=5$}

BHs in a 5D (4+1) spacetime is considered in Ref. \cite{Anchordoqui:2025opy}, with primary focus on primordial BHs. Recalling the choice of metric being Eq. (\ref{tangh}), which across this work we have chosen $p=2$ in $n=4$, we briefly consider the $n=5$ Tangherlini-like black brane metric, for both point-like $p=0$ and surface-like $p=2$ BHs. 
\begin{itemize}
\item For point-like:
\begin{equation}
ds^2=-\left(1-\frac{r_S^2}{r_4^2}\right)dt^2+\left(1- \frac{r_S^2}{r_4^2}\right)^{-1} dr_4^2+r_4^2d\Omega^2_{3},
\end{equation}
where $r_4=\sqrt{x^2+y^2+z^2+\zeta^2}$ is the radial coordinate in four spatial dimensions. The $d\Omega^2_3=d\psi^2+\sin^2\psi d\Omega^2_2$ element for a 3-sphere \cite{Landau:1988} ($\psi\in[0,~\pi]$) is obtained. In the $n$-dimensional spacetime \cite{Matyjasek:2015qra, Nasseri:2002uy}, there are higher-dimensional definitions for the Schwarzschild radius and the gravitational constant:
\begin{equation}
r_S^{n-3}=\left(\frac{16\pi G_nM}{(n-2)\Omega_{n-2}} \right)\quad\mathrm{and}\quad G_n=\frac{4\pi V_{n-4}}{\Omega_{n-2}}G,
\end{equation}
where $V_{n-4}$ is the volume of the extra spatial dimensions, and $\Omega_{n-2}$ is the area of a $(n-2)$-unit sphere. The case of $n=4$ recovers $r_S=2GM$ with $G$ being the familiar Newton-Cavendish constant for gravity. For $n=5$, $r_S^2=32GMV_1/(3\pi^2)$, which gives the Schwarzschild potential the inverse-square law, and the $G_5$ constant a larger strength, in the higher dimension. \\
\item For surface-like:
\begin{eqnarray} 
&ds^2=-dt^2+ dr_4^2+r_4^2d\Omega^2_3\\\nonumber
&+r_4^2d\phi^2+\sin^2\psi\Bigg(\sin^2\theta \left(dr_4^2-r_4^2 d\theta^2\right)+2r_4\sin\theta\cos\theta d\theta dr_4 \Bigg)\\\nonumber
& +r_4^2d\psi^2(\cos^2\psi \sin^2\theta-1)\\\nonumber
&+r_4\sin\theta \sin(2\psi) d\psi \Big(r_4\cos\theta d\theta+\sin\theta dr_4\Big),
\end{eqnarray}
where the flat 4+1 metric is accompanied by a subleading metric that contains higher-dimensional angular effects. Setting aside the subleading metric as though it were the perturbation part of the $g+h$ split, the background metric is therefore flat. This may be instrumental to the bulk-viscous relation via Eq. (\ref{bulky}), which assumes flatness everywhere.
\end{itemize}

\section{Conclusion} \label{sect:concl}

In this work, we analyzed a Schwarzschild BH as though it contained an internal mirror-bulk underneath its horizon. Under the choice of metric being the Tangherlini-like $p$-brane metric, we specifically considered the metric in physical $n=4$ and brane $p=2$ dimensions, approximating the BH horizon as a closed spherical surface in our reality. In following the holographic principle and implying AdS/CFT, we face the consequence of analyzing the holographic "mirror-bulk" within the BH, whose dynamics are mirrored by the surface-dwelling Hawking particle states.

Under our choice of metric, the BH interior behaves as a linear well, with a natural resting point along the BH horizon rather than the geometric center. From the outside, under the quintessential Schwarzschild metric, the resting point of infalling objects is also the BH horizon, perserving what is well-understood. And so, the choice of metric used across this work is presented as a piecewise metric, given as Eq. (\ref{newmet}).

In taking on the variational approach to the Einstein field equations, where the geometry of a metric Ansatz approximates the effective energy content, the energy density profile of $T_{00}$ was found for both inside and outside the BH, allowing the metric Ansatz to be Eq. (\ref{newmet}). Along with internal energy (density) (Eqs. [\ref{t00}] and [\ref{ein}]), the radial pressure $T_{11}$ was found as Eq. (\ref{t11}), and with it how one relates energy density and pressure into a BH ``equation of state" in the context of BH thermodynamics.

In the future, we want to explore the BH mirror-bulk in-depth beyond its introduction here. Specifically, it is the hope to explore the mirror-bulk material on its own, which implies how microphysical interactions occur in a linear-well background imposed by Eq. (\ref{newmet}). It is also of interest to introduce (slow) static rotations to the model, which extends the mirror-bulk analysis from Schwarzschild BHs to slow-Kerr BHs \cite{Page:1976ki}.

\section*{Acknowledgments}

I thank Percy Martinez for having inspired this mirror-bulk model for astrophysical Schwarzschild BHs. This work and the scope therein is motivated by the mission statements of working group WG3 of COST-Action CA24139 -- Superfluid Condensates in Astrophysics and Laboratory Experiments (SCALES).

\vfill

\section*{Statement Declarations}

\subsection*{Conflict of Interest}
The author declares no conflicts of interest.

\subsection*{Data Access Statement}
As a theoretical study, this work generates no original data. 

\subsection*{Ethics Statement}
No ethical issues arise, as no test subjects are involved. This paper adheres to academic integrity.

\subsection*{Funding Statement}
This work received no funding.





\end{document}